\documentclass[runningheads,a4paper]{llncs}

\usepackage{amssymb}
\usepackage{graphicx, booktabs}

\usepackage{url}
\urldef{\mailsa}\path|{alfred.hofmann, ursula.barth, ingrid.haas, frank.holzwarth,|
\urldef{\mailsb}\path|anna.kramer, leonie.kunz, christine.reiss, nicole.sator,|
\urldef{\mailsc}\path|erika.siebert-cole, peter.strasser, lncs}@springer.com|

\begin{document}

\mainmatter  

\title{``To Share or not to Share'' in Client-Side Encrypted Clouds}

\titlerunning{``To Share or not to Share'' in Client-Side Encrypted Clouds}

%
%
\author{Duane C. Wilson \and Giuseppe Ateniese}
\authorrunning{``To Share or not to Share'' in Client-Side Encrypted Clouds}

\institute{Johns Hopkins University, Computer Science  Department\\Information Security Institute, Baltimore, Maryland\\
}

%
%

\toctitle{Lecture Notes in Computer Science}
\tocauthor{Authors' Instructions}
\maketitle

\begin{abstract}
With the advent of cloud computing, a number of cloud providers have arisen to provide Storage-as-a-Service (SaaS) offerings to both regular consumers and business organizations.  SaaS (different than Software-as-a-Service in this context) refers to an architectural model in which a cloud provider provides digital storage on their own infrastructure.  Three models exist amongst SaaS providers for protecting the confidentiality of data stored in the cloud: 1) no encryption (data is stored in plain text), 2) server-side encryption (data is encrypted once uploaded), and 3) client-side encryption (data is encrypted prior to upload).  Through a combination of a Network and Source Code Analysis, this paper seeks to identify weaknesses in the third model, as it claims to offer 100\% user data confidentiality throughout all data transactions.  The weaknesses we uncovered primarily center around the fact that the cloud providers we evaluated (Wuala, Tresorit, and Spider Oak) were each operating in a Certificate Authority capacity to facilitate data sharing.  In this capacity, they assume the role of both certificate issuer and certificate authorizer as denoted in a Public-Key Infrastructure (PKI) scheme - which gives them the ability to view user data contradicting their claims of 100\% data confidentiality.  We have collated our analysis and findings in this paper and explore some potential solutions to address these weaknesses in these sharing methods. The solutions proposed are a combination of best practices associated with the use of PKI and other cryptographic primitives generally accepted for protecting the confidentiality of shared information. 
\end{abstract}

\section{Introduction}

Cloud computing is a model for enabling convenient, on-demand network access to a shared pool of configurable computing resources (e.g., networks, servers, storage, applications, and services) that can be rapidly provisioned and released with minimal management effort or service provider interaction. This cloud model promotes availability and is composed of five essential characteristics (On-demand self-service, Broad network access, Resource pooling, Rapid elasticity, Measured Service); three service models (Software as a Service, Platform as a Service, Cloud Infrastructure as a Service); and four deployment models (Private cloud, Community cloud, Public cloud, Hybrid cloud)~\cite{ref1}.

With the advent of cloud computing, a number of Cloud Storage Providers (CSPs) have arisen to provide Storage-as-a-Service (SaaS) offerings to both regular consumers and business organizations.  SaaS refers to an architectural model in which a cloud provider provides digital storage on their own infrastructure~\cite{ref2}. Three models exist amongst SaaS providers for protecting the confidentiality of data stored in the cloud: 1) no encryption (data is stored in plain text), 2) server-side encryption (data is encrypted once uploaded), and 3) client-side encryption (data is encrypted prior to upload).  Dropbox is the most popular version of a cloud storage provider that adheres to the first confidentiality model.  In this paper we examine secure alternatives to  Dropbox that provide client-side encryption.  Our primary motivation is based on consistent claims made by CSPs that guarantee the confidentiality of data stored in the cloud.  The major claims are as follows:\\
\begin{itemize}
\item``No one unauthorized not even the cloud storage provider can access the files.'' \cite{ref3}
\item``Our 'zero-knowledge' privacy environment ensures we can never see your data. Not our staff. Not government. Not anyone.'' \cite{ref4}
\item``Contrary to other solutions, no storage provider or network administrator, no unauthorized hacker, not even we can read your files.'' \cite{ref5}
\end{itemize}

These principles of confidentiality hold true in use cases where data is not shared with other cloud users or with entities outside of the cloud storage environment (e.g., non-members).  In the evaluated SaaS environments, data sharing is accomplished in three ways: Web Link, Folder, or Group, the latter two being the focus of this research. Through our analysis we discovered that for each data sharing mechanism, there are inherent weaknesses that can expose user data to the cloud provider which directly contradicts the aforementioned CSP claims. What would prompt a Cloud Provider to compromise the trust of its users in this way?

\begin{enumerate}
\item  \textbf{National Security}: In the interest of National Security, governments will often collect citizen data for the purposes of confirming a threat.  This collection typically consists of activities such as wire tapping, data harvesting, and other forms of information collection.  Data Harvesting (usually associated with Cloud-Storage attacks) refers to the collection of disparate data from a homogeneuous location.  This approach is advantageous for a Government seeking data points from multiple entities such as is the case in a co-resident Cloud Storage environment.   
\item  \textbf{Oppressive Government}: Under certain government regimes, there may exist situations in which a Government might ``force'' a Cloud Provider to comply with new or existing disclosure regulations.  Reasons for this may include: 1) Company Sanctions, 2) Periodic Evaluations, or 3) to Limit Monopoly business practices.
\item  \textbf{Data Leakage Confirmation}  As is common in the ``Networked'' world, data that is accessible via the Internet is subject to data breaches.  The Data Breach Notification Laws require states to report the occurrence of company and state-related data breaches.  In order to confirm and subsequently comply with these laws, Cloud Providers could potentially need to access user data in plain text form. 
\end{enumerate} 

Our primary contributions are as follows:
\begin{itemize}
\item We present an overview of the various sharing scenarios employed by each evaluated CSP 
\item We highlight the weaknesses  found in each CSP sharing scenario.  Our research focus is on exposing the weaknesses in private Group and Folder sharing scenarios.  
\item We describe how an attack against private Group or Folder sharing functions could work in practice.
\item We provide evidence of Certificate Authority functionality via network traffic and source code analysis.
\item We reverse engineer various CSPs code to reveal evidence that substantiates our claims that user data is not 100\% safe from being read and/or manipulated by the CSP.  
\item We provide suggestions for addressing the inherent weaknesses discovered in the design of the CSP sharing functions. 
\end{itemize}The rest of this paper is organized as follows:  Section 2 discusses the work related to our analysis area, Section 3 provides an overview of the CSPs we evaluated and the sharing scenarios they support, Section 4 discusses the attack and associated threat model, Section 5 provides a detailed review of our analysis methodology and associated results, and Sections 6 concludes the paper and provides areas for future work.

\section{Related Work}

This section puts our analysis in perspective by examining preceding work that relates to the analysis of Cloud Storage Providers (CSPs).  Prior work falls into three specific categories: 1) General Analysis of CSP security capabilities, 2) Analysis of the Design an Implementation of CSPs, and 3) Analysis of the Weaknesses associated with CSPs.
  
In \cite{ref6}, Borgmann and Waidner of the Fraunhofer Institute for Secure Information Technology, studied the security mechanisms of seven Cloud Storage Services: CloudMe, CrashPlan, Dropbox, Mozy, TeamDrive, Ubuntu One, and Wuala.  The study focused on the following security requirements: Registration and Login, Transport Security, Encryption, Secure File Sharing, and Secure Deduplication.  Similar to our approach, they examine several aspects of the CSP's file sharing mechanisms for security flaws.  Specifically, they highlight the fact that if client-side encryption is used, sharing should not weaken the security level.  In particular, the CSP should not be able to read the shared files.  The scope of their work covered sharing with other subscribers of the same service, sharing files with a closed group of non-subscribers, and sharing files with everybody.  As we also discovered in our research, sharing via secret web link (e.g., closed group of non-subscribers) or making data public reveals shared data to the CSP.  Their analysis, however, did not yield the fact that sharing files with subscribers could also result in a breach of confidentiality with the CSP due to the CSP acting in a Certificate Authority (CA) capacity.  This is proven in our analysis and can serve as a viable extension to their work.
  
Mager et al. in \cite{ref7} examine the design and implementation of an online backup and file sharing system called Wuala.  The goals of their work consist of four primary items: Characterization of the Infrastructure, Understanding the Data Placement Methodology, Identifying the Coding Techniques relating to Data Availability (accessibility of files at any time) and Durability (ensuring files are never lost), and the Determination of the Data Transport Protocol used.  The scope of their evaluation does not necessarily include security considerations, although the data structure employed by Wuala for sharing data is mentioned as well as the type of encryption used for performing client-side encryption.  Overall, this research provided useful insight into methods employed by the CSP to facilitate client usage of their infrastructure, however, it does not examine the infrastructure or design in detail for security weaknesses.  As it pertains to their goals, our work differs in that it looks at the CSP infrastructure security features for consistency with user expectations.  We focus on the data sharing (not placement) methodology employed by the CSPs.  The coding techniques we identified relate to the discovery of evidence that the CSP is operating in a CA capacity during sharing transactions.
  
Finally in \cite{ref8}, Kholia and Wegrzyn analyze the Dropbox cloud-based file storage service from a security perspective.  Their research presents novel techniques to reverse engineer frozen Python applications (to include Dropbox). They specifically describe a method to bypass Dropbox's two factor authentication and hijack Dropbox accounts. Additionally, they introduce generic approaches to intercept SSL data using code injection techniques and monkey patching.  This work is consistent with our analysis of CSPs for major security flaws as it results in the exposure of private user data and enables the CSP to have access to shared user data.   

\section{Cloud Storage Provider Overview}
This section provides an overview of the Cloud Storage Providers (CSPs) we evaluated and the `confidentiality model' they each employ to protect the confidentiality of user data.  We define a confidentiality model as the method employed by the CSP to protect the confidentiality of users' stored data.  According to \cite{ref9,ref10}, there are several CSPs that offer client-side encryption.  For our analysis, we focused on Wuala, Spider Oak, and Tresorit. Wuala encryption is performed with AES256 prior to files being uploaded.  Encryption comprehensively includes not only file content, but also: file names, preview images, folders and metadata. An RSA2048 key is used for signatures and key exchange when folders are shared while SHA-256 is used for data integrity.  Spider Oak can be used to share and back up files.  Data is encrypted on a user computer with AES256 in CFB mode and HMAC-SHA256.  As mentioned previously, the company claims to have no knowledge of what data is stored in their servers or user passwords. Their software works in smart phones, the Linux operating system, and Windows~\cite{ref9}. Lastly, Tresorit is a Hungarian-based company that uses AES256 to encrypt data before uploading it to the cloud. The company is offering \$10,000.00 (US) to anyone who can break their security software.  Similar to Spider Oak, data can be accessed via smart phone or desktop computer.  Each CSP's confidentiality model is discussed in the following subsections.  All information in this section has been adapted from the CSP's website and/or documentation unless otherwise specified.    
 
Table 1 summarizes the confidentiality model employed by each CSP based on each data sharing scenario provided.  It also presents an overview of each sharing scenario according to the level of effort required to 'exploit' the scenario (Trivial or Non-Trivial).  Our research focuses on the sharing scenarios that the CSPs highlight as offering 100\% data confidentiality (i.e., Private Group and Private Folder sharing).  ``N/A" represents a feature that is not implemented or has no documentation to support the scenario.

\begin{table} 
\centering 
\begin{tabular}{l c c c c c c} 
\toprule 
& \multicolumn{5}{c}{ Sharing Scenario} \\ 
\cmidrule(l){2-7} 
& Public & Private & Public & Private & Public & Private\\ 
CSP & Web Link & Web Link & Folder & Folder & Group & Group\\ 
\midrule 
Wuala & Public & Private & Public & Public Key & Public & Public Key\\ 
& URL & URL & Folder & Encryption & Group & Encryption\\ 
 &  & & & & \\ 
 &  & & & & \\ 
Spider Oak & Public & Public URL & N/A & N/A & N/A & N/A\\ 
& URL & with Passwd & & & &\\ 
 & & & & & \\ 
 &  & & & &\\ 
Tresorit & N/A & Encrypted & N/A & RSA & N/A & ICE\\ 
 & & Link & & or & & Protocol\\ 
 & & & & TGDH & \\ 
\midrule 
\midrule 
Difficulty & Trivial & Trivial & Trivial & \textbf{Non-Trivial} & Trivial & \textbf{Non-Trivial}\\ 
\bottomrule 
\end{tabular}
\caption{Trivial vs. Non-Trivial Sharing Scenarios} 
\label{tab:template} 
\end{table}

As noted above for all CSPs, when data is shared via Private web link, the CSP requires the transmission of some sensitive data in order to process the user's request.  Similarly, each of the Public Sharing Scenarios (i.e., Web Link, Group, Folder) require implicit trust of the CSP in order to make the data public.  As a result, these scenarios were discussed in this section for completeness but fall outside of the scope of our research due to this requirement of trust.   

\section{Threat Model}
In this section we provide a threat model which describes how an attack against a Cloud Storage Provider (CSP) offering client-side encryption could potentially be executed.  
Rating the threats we identify in this section are outside of the scope of our research.  We make the following assumptions with regards to the threat model:
\begin{enumerate}
\item The CSP client is trusted and has not been previously modified by a malicious insider or outsider.  In the case 	of a modified client, users could easily be redirected to a rogue server upon client startup and be forced to perform all sharing transactions through this server.  
\item The CSP server is trusted and has not been compromised.  Similar to the case of a rogue client, a rogue server would allow an adversary access to user certificates and gain access to decryption keys.  
\item Other CSP members can be trusted.  When searching for an individual user in the CSP database, it is assumed that the users identified are legitimately who they appear to be.    
\item The User certificates issued cannot be trusted because they are issued by the CSP.
\item Public or Private Web Link Sharing scenarios (as mentioned above) require users to reveal some form of sensitive information to the CSPs to enable decryption in a browser environment.  Similarly, Public Groups and Folders are accessible by all members and pose no threats to users.
\end{enumerate} Our analysis consists primarily of network traffic analysis and reverse-engineering the source code of each CSP client through decompilation or disassembly.  We selected this course of analysis due to fact that most of the CSP client code used some form of code obfuscation --- making the replication of a rogue client challenging.   Additionally, we could not produce counterfeit certificates `on behalf of' other users, because we do not possess the signing key of the CSPs we analyzed.  This signing key would be required to produce certificates that CSP client applications would trust.  We discuss below a scenario that would enable a Cloud Provider to maliciously access a user's data. 

\begin{enumerate}
\item User A signs up for Cloud Account
\item User A initiates sharing request with User B
\item Cloud client returns User B's contact information to User A
\item Cloud Provider substitutes its Public Key for User B
\item Without User A's knowledge his/her data is encrypted with the Public Key of the Cloud Provider
\item Cloud Provider is able to decrypt User A's data and view its contents
\item Cloud Provider then re-encrypts data with User B's Public Key and Cloud Client sends sharing request to User B
\item User B decrypts the data sent by User A w/o knowledge of the above-stated attack.
\end{enumerate}

This scenario simply show that because the CSPs are operating as Certificate Authorities, certificate manipulation, spoofing, or substitution of any kind can be accomplished. It is also important to emphasize that a malicious CSP cannot be detected since it can perform a standard \emph{man-in-the-middle} attack in which the encrypted information flow from the sender is first decrypted and then re-encrypted on the fly under the recipient's public key (and vice-versa).  In essence, the CSP is able to spoof the identity of any user through the use of certificates.  In the next section, we show evidence to support our claim that each evaluated CSP is serving in a CA capacity --- making all users susceptible to the aforementioned attacks.

\section{Analysis Methodology and Results}
In this section we provide a detailed review of our analysis results.  It is important to note that although our results do show that each Cloud Storage Provider is in some capacity operating as a Certificate Authority (i.e., could issue fake certificates to its users) - we did not find any evidence to support any claims of misuse by the Providers themselves (i.e., they are not actively exploiting users via this attack).   Our analysis consisted of three phases examining the sharing functions employed by each CSP: 1) Network Traffic Analysis (using Wireshark \cite{ref12} and Fiddler Web Proxy Debugger \cite{ref13}), 2) Source Code Decompilation (using AndroChef Java decompiler \cite{ref14} and Boomerang C++ decompiler), and 3) Source Code Disassembly (using HopperApp \cite{ref15} and Synalyze \cite{ref16}).  Evidence of CA functionality validates our assertion that when data is shared, it is possible for the CSP to view user data  which contradicts their 100\% data confidentially claims.   Evidence of CA functionality includes (but is not limited to the following): Certificates (i.e., Root or Intermediary), Certificate Issuance, Certificate Validation, Certificate Revocation, Certificate Authentication, Certificate Renewal, Certificate Registration, Obtaining Certificates, Encryption/Decryption, and Digital Signature (signing or verification).  The processes for performing the network traffic, source code decompilation, and source code disassembly analyses are discussed below.   

A CA is a Trusted Third Party (TTP) organization or company that issues digital certificates used to create digital signatures and public-private key pairs. The role of the CA in this process is to guarantee that the individual granted the unique certificate is, in fact, who he or she claims to be. CAs are a critical component in data security and electronic commerce because they guarantee that the two parties exchanging information are really who they claim to be~\cite{ref17}. Serving as a CA, the CSPs would potentially be able to issue counterfeit certificates to users -- pretending to be a legitimate entity that data is shared with.  The code we identified falls into five categories: 1) Code Libraries, Certificate Files, Certificate Operations, Keystore Files/Definitions, and Peripheral Cryptographic Functions.  Lastly, we focused on disassembling code that could not be reproduced in its original form through Decompilation.  The code samples we extracted were only those that could potentially represent CA functionality.  We primarily analyzed the main binary files ( \textit{.app}, \textit{.exe} files) associated with each CSP.    

\subsection{Wuala Analysis}
We started our analysis of Wuala by examining the network traffic that was generated between the client and server during sharing transactions and extracting any evidence of CA functionality.  Through our analysis, we discovered evidence of a Wuala CA certificate which was used to confirm the validity of the other certificates below it in the certificate chain.  A root certificate is either an unsigned public key certificate or a self-signed certificate that identifies the Root Certificate Authority (CA).   It is the top-most certificate of the tree, the private key of which is used to ''sign'' other certificates. All certificates immediately below the root certificate inherit the trustworthiness of the root certificate~\cite{ref18}.

Secondly, during disassembly of the Wuala binaries, we identified multiple references to a 'WualaCACerts' file  which contains several certificates that the Wuala client uses during its execution  to include the Wuala CA file --- mentioned previously.  The reference below is to the bouncy castle (Java Crypto Library) implementation of a Keystore which also contains individual certificate entries.  The contents of this particular file is discussed in the decompilation section below. Lastly, during the decompilation phase, we discovered that Wuala uses a combination of Name Obfuscation (e.g., changing method and variable names while maintaining the functionality).  To overcome this, we performed searches across the entire codebase for files that fit into the aforementioned categories.  The primary results of the Wuala decompilation process yielded the WualaCACerts file.  This file contains 4 certificate entries: \textit{wualaadminca}, \textit{wualaserverca}, \textit{wualaca}, and \textit{wualaclientca}.  Each certificate is named to denote the function it serves within the operation of the Wuala application (i.e., administrative, client, server, and CA).  Each certificate entry has an 'Extension' section with an ObjectID and a specification of what the key (within the certificate file is used for).  In each of these, the key is used for Certificate Signing.  According to IBM's \textit{Key usage extensions and extended key Usage} page, the Certificate Signing key usage extension is to be used when the subject public key is used to verify a signature on certificates. This extension can be used only in CA certificates~\cite{ref19}. We summarize the contents of the Wuala CA certificate below: 

\begin{verbatim}
Alias name: wualaca, Creation date: Jan 6, 2012
Entry type: trustedCertEntry, Owner: CN=Wuala CA, OU=Wuala, 
EMAILADDRESS=cert@wuala.com, Issuer: CN=Wuala CA, OU=Wuala, O=LaCie 
EMAILADDRESS=cert@wuala.com, Signature algorithm name: SHA1withRSA
\end{verbatim}
As indicated by the algorithms identified in the certificates as well as the key usage extension, this certificate is used for creating and verifying digital signatures.  However, the same certificates could be issued by Wuala for performing other PKI-related functions  to include both encryption and decryption of user data or the creation of private groups. 

\subsection{Spider Oak Analysis}
In the case of Spider Oak, the disassembly of its binary did not yield results to substantiate our claims, thus we focused primarily on the results of decompilation and network traffic analysis in this section.  During the code analysis of Spider Oak, we uncovered a 'Public Key' folder containing DSA files, RSA files, a public key file, and other crypto-related files.  These files were each disassembled and evaluated for evidentiary purposes.  The `RSA.pyc' file is presented below as a representative sample.  It shows a call to (or definition of) a `generate' function used to generate an RSA public/private key pair programmatically.  There is a similar file for generating a DSA key pair  which we excluded.  The generation of certificates in a programmatic function confirms that users are not allowed to use their own certificates for cryptographic purposes  which can allow Spider Oak to generate certificates 'on behalf' of legitimate users.  As a result, a user could be tricked into sharing data with Spider Oak instead of the intended user.

\begin{verbatim}
db "generate(bits:int, randfunc :callable, progress_func:callable)    
Generate an RSA key of length 'bits',  using 'randfunc' to get 
random data and 'progress_func', if present, to display the 
progress of the key generation. l\x02"
\end{verbatim} 
Unlike Wuala, Spider Oak does not appear to use any form of code obfuscation.  The files we examined contained the .pyc extension which is the equivalent of a Java Byte Code file for the Python programming language.  To analyze these files, we had to decompile them to the original Python Source code.  Similar to Wuala, Spider Oak uses a file called `certs.pyc' to store its certificates.  This file contained 5 certificate entries: Equifax Secure CA, GeoTrust Global CA (2), RapidSSL CA, and the Spider Oak Root CA.  Contents of the Spider Oak Root CA certificate are as follows:\\
\begin{verbatim}
Data:  Version: 1 (0x0), Serial Number: ea:14:d7:ad:6a:cf:dc:35     
Signature Algorithm: sha1WithRSAEncryption         
Issuer: emailAddress=ssl@spideroak.com, organizationName=SpiderOak             
\end{verbatim}

This analysis phase revealed a self-signed root certificate from Spider Oak.  A self-signed certificate is an identity certificate that is signed by the same identity it certifies.  As mentioned above, a typical PKI scheme requires certificates to be signed by a Trusted Third Party (TTP) for verification purposes.  Self-signed certificates cannot (by nature) be revoked, which may allow an attacker who has already gained access to monitor and inject data into a connection to spoof an identity if a private key has been compromised~\cite{ref20}. In this case, the attacker could potentially be Spider Oak (e.g., spoofing the identity of another user and gaining access to user data).   

Lastly, examining the network traffic between the Spider Oak client and server produced the same root certificate identified during Decompilation.  As a root certifier, Spider Oak is also responsible for validating any certificate in its trusted certificate chain -- to include Spider Oak client certificates issued to users.  Finding this type of certificate also demonstrates the fact that as a certificate issuer, Spider Oak can issue user certificates that spoof the identity of another user  resulting in a data confidentiality breach when data is shared with another Spider Oak user.  Similar to the Wuala example (discussed above), the Spider Oak Client certificate is also validated by its respective Spider Oak Root CA.

In a recent article, Eric Snowden urges consumers to adopt more secure file storage systems which are less susceptible to government surveillance - mentioning Spider Oak specicially \cite{snowden}.  However, he fails to disclose (or realize) the fact that data confidentiality could be breached if data is shared (as shown above) within these same environments.  Additionally, Spider Oak was interviewed by Senior Writer Brian Butler of Network World concerning some of our findings \cite{networkworld}.  The company claims that the features we evaluated fall outside of the scope of features that they have implemented \cite{spideroak}, however, we assert that in order to share encrypted data - some form of Public Key Encryption must be used.  As a result, Spider Oak users are still subject to this type of flaw.  
\subsection{Tresorit Analysis}
During the Decompilation of Tresorit, we did not identify any independent certificate files  as was the case with  Wuala and Spider Oak.  According to Tresorit's documentation, a roaming profile file contains both the SSL/TLS client certificate and the associated private key required to communicate with Tresorit's servers, and agreement certificates and private keys to securely access and decrypt the contents of a Tresor.  Hence, there were not any specific certificate entries to examine during this phase.  

Disassembly of the Tresorit binary contained a number of references to certificate operations, files, and uses with application.  The entry below shows the loading of a certificate in the traditional X.509 format.  This confirms that X.509 certifications are utilized in the Tresorit client application for cryptographic functions, however, does not fully substantiate the fact that Tresorit is operating in a Certificate Authority (CA) capacity.
\begin{verbatim}
db "y_OBJ", 0, db  "X509_get_pubkey_parameters", 0
db "X509_load_cert_crl_file", 0, db "X509_load_cert_file", 0
db "X509_load_crl_file", 0
\end{verbatim}In the network analysis of Tresorit, we identified a certificate issued by Tresorit --- the Tresorit User CA certificate.  This certificate differs from the Wuala and Spider Oak certificates in that it is validated by the StartCom CA  which would imply that they are using a TTP to ensure 100\% data confidentiality during sharing transactions.  However, the fact that this certificate was issued by Tresorit verifies that they also can view user data whenever the keys are used for encryption/decryption purposes.  Additionally, the fact that users are neither allowed to use their own certificates or enter credentials to generate their own certificates upon signing up with any of the three CSPs gives credence to our claims that counterfeit certificates could be issued.  This would lead users to implicitly trust that other users are who they say they are (i.e., the CSP is not acting on their behalf).  In the case of Tresorit, StartCom CA is serving as the Root CA with Tresorit serving as an Intermediate CA.  This differs from both Wuala and Tresorit, however, is still indicative of Tresorit operating as a CA in some capacity.

\section{Conclusions and Future Work}
In this paper we examined three major Cloud Storage Providers (CSPs) for weaknesses associated with their sharing functions.  Each CSP claims to offer 100\% user data confidentiality through all data transactions executed within their respective cloud instances.  We discovered that these principles of confidentiality hold true in use cases where data is not shared with other cloud users or with entities outside of the cloud storage environment (e.g., non-members).  We presented several scenarios in which data confidentiality could be breached to include: when data is shared via Public Web Link, when data is shared via Private Web Link, when CSP is accessed via Web Browser (i.e., web access), and when data is shared via Group or Private folder (due to the CSP being the issuer of the certificates used for cryptographic operations -- hence could allocate counterfeit  certificates to users).  We provide evidence in the form of Network Traffic and Source Code analysis for each CSP to substantiate these claims.

To address the weaknesses discussed in the previous sections, users should be issued standard certificates via a Trusted Third Party  which is typical in a traditional Public/Private Key implementation or should be allowed to use their own certificates (e.g., PGP). Alternatively, out-of-band verification mechanisms should be used such as those provided by several implementations of off-the-record (OTR) messaging [21]. Similarly, Tresorit makes use of ICE --- a proprietary protocol that is designed to work in a semi-trusted environment. It does not rely solely on the trustworthiness of the certificate authority issuing the users' certificates, but also an invitation secret and optional pairing password. Lastly, to enhance the security associated with sharing data via Private Web Link, it is possible to require password authentication prior to listing shared files.  Spider Oak employs such a strategy, although the CSP would still be required to verify that the password was entered correctly.  In the future, we plan to extend this work by examining alternative methods of secure file sharing that do not rely on a CA for identity-verification - which will consequently overcome a number of weaknesses we have identified in this research.

\end{document}